\documentclass[showpacs,prl,preprint]{revtex4-1}
\usepackage{latexsym}
\usepackage{amsfonts}
\usepackage{amssymb}
\usepackage{amsmath}
\usepackage{graphicx}

\begin{document}
\title{Fingerprints of Inelastic Transport at the Surface of Topological Insulator Bi$_{2}$Se$_{3}$: Role of Electron-Phonon Coupling}
\author{M. V. Costache,$^1$ I. Neumann,$^{1,2}$ J.  F. Sierra,$^1$ V. Marinova,$^3$ M. M. Gospodinov,$^4$ S. Roche,$^{1,5}$ and S. O. Valenzuela$^{1,2,5}$}
\affiliation{$^1$ICN2 - Institut Catala de Nanociencia i Nanotecnologia, Campus UAB, Bellaterra, 08193 Barcelona, Spain}
\affiliation{$^2$Universitat Auton\'oma de Barcelona, Bellaterra, 08193 Barcelona, Spain}
\affiliation{$^3$Institute of Optical Materials and Technologies, Bulgarian Academy of Science, Sofia 1113, Bulgaria}
\affiliation{$^4$Institute of Solid State Physics, Bulgarian Academy of Sciences, 72 Tzarigradsko Chaussee blvd, 1784 Sofia, Bulgaria}
\affiliation{$^5$ICREA - Instituci\'o Catalana de Recerca i Estudis Avan\c{c}ats, 08010 Barcelona, Spain}
\email{mcostache@icn.cat} \email{SOV@icrea.cat}
\date{\today}

\begin{abstract}
We report on electric-field and temperature dependent transport measurements in exfoliated thin crystals of Bi$_{2}$Se$_{3}$ topological insulator. At low temperatures ($< 50$ K) and when the chemical potential lies inside the bulk gap, the crystal resistivity is strongly temperature dependent, reflecting inelastic scattering due to the thermal activation of optical phonons. A linear increase of the current with voltage is obtained up to a threshold value at which current saturation takes place. We show that the activated behavior, the voltage threshold and the saturation current can all be quantitatively explained by considering a single optical phonon mode with energy $\hbar \Omega \approx 8$ meV. This phonon mode strongly interacts with the surface states of the material and represents the dominant source of scattering at the surface at high electric fields.
\end{abstract}
\maketitle


The observation of surface states protected by time inversion invariance in the Bi$_{2}$Se$_{3}$ family of materials \cite{hasan10,moore10,zhang11,brune11,xia09,zhang09}, have triggered intense research because of the possibility of developing topological insulator (TI) devices at room temperature. However, the properties limiting charge transport, which is key for electronic applications, are still not known in detail. Elastic scattering by disorder imposes a limit to the conductivity of the surface states at low temperatures, nevertheless, as disorder is reduced, the limit at finite temperatures will be ultimately set by the intrinsic electron-phonon (\textit{e-ph}) coupling. In recent theoretical work a strong \textit{e-ph} coupling in Bi$_{2}$Te$_{3}$ was obtained \cite{egger11,giraud12}. The results are in agreement with temperature-dependent angle-resolved photoemission spectroscopy (ARPES) measurements in Bi$_{2}$Se$_{3}$ \cite{hofmann11,zhou13}. ARPES studies target the electronic structure and integrate the \textit{e-ph} interaction over all phonon modes. The main dispersive surface optical-phonon branch with an energy $\hbar \Omega$ of about 6 to 8 meV at the $\overline{\Gamma}$ point was identified using helium-beam surface scattering, which also showed the absence of Rayleigh phonons \cite{batanouny11}. The \textit{e-ph} coupling constant $\lambda = 0.43$ for this branch was found to be larger than any of the integrated values that are reported with ARPES \cite{batanouny12}. Such results are further supported by magneto-optical \cite{basov10} and Fourier-transform interferometry \cite{dipietro13} which found an optical phonon mode at $\hbar \Omega = 7.6$ meV.

Experimental investigations therefore suggest an anomalously large \textit{e-ph} coupling for a specific surface-phonon branch, which could be readily observable in electrical transport measurements, \cite{wu13} in particular because surface states in Bi$_{2}$Se$_{3}$ carry a large fraction of the current flowing in thin crystals \cite{peng09,steinberg10,ong11,sacepe11,xiu11} and films \cite{steinberg11,kandala13}. Additionally, the calculated bulk phonon dispersion curves projected into the surface Brillouin zone also show an optical phonon in the same energy range \cite{batanouny11}. However, no fingerprint of an optical phonon mode on inelastic transport has been reported to date.

In this Letter, by performing temperature and voltage dependent transport measurements in Bi$_{2}$Se$_{3}$ thin crystals, a strong electron-phonon mediated inelastic backscattering phenomenon is unveiled. For temperatures $T < \hbar \Omega/k_B \approx 90$ K ($k_B$ is the Boltzmann constant), the resistance is highly nonlinear and is consistent with a thermally-activated behavior dictated by an optical phonon with $\hbar\Omega\simeq 8$ meV, as identified in spectroscopic experiments \cite{batanouny11, batanouny12, basov10,dipietro13}. Moreover, the onset of suppression of the conductance for $eV \gtrsim 10$ meV, in conjunction with current saturation as reported in graphene \cite{meric08,barreiro09,Foa2006}, further confirm the influence of a $\hbar\Omega\simeq 8$ meV mode.

The devices were fabricated with single crystals mechanically exfoliated onto a highly-doped Si substrate with 280 nm or 440 nm of thermally grown SiO$_{2}$, followed by e-beam lithography, metal deposition, and lift-off. The inset of Fig. 1(c) shows an optical microscope image of a typical device used in the present study. The chosen crystals have elongated shape to obtain a homogeneous current flow. The distances between the inner and outer electrodes are about 0.3 - 1 $\mu$m and 2 - 3 $\mu$m, respectively. Four-probe transport measurements down to 4.2 K were performed on six different devices in vacuum, yielding similar results (Table SI \cite{SI}). Here we present representative data for two of them with $t=20$ nm (device D1) and $t = 30$ nm (device D2).

Figures 1(a) and (b) show the square resistance $R$ as a function of gate voltage $V_{g}$ for devices D1 and D2, respectively. When sweeping $V_{g}$ from positive to negative values, an increase in $R$ is observed, which is due to the $n$-type doping commonly observed in Bi$_{2}$Se$_{3}$ crystals. Figure 1(c) shows $R$ versus temperature $T$ for device D1 at three gate voltages (+30, -60, and -100 V) and $I = 100$ nA. For $V_{g}$ = 30 V, $R$ is weakly temperature dependent, presumably because the chemical potential $\mu$ lies inside the conduction band and disorder scattering is dominant \cite{ong11}. In contrast, the temperature dependence of $R$ becomes highly nonlinear when $\mu$ is shifted towards the material bulk gap, where the surface transport is expected to dominate \cite{steinberg10,ong11}. This is evident for $V_g$ = -60 and -100 V and from weak localization measurements \cite{ong11,steinberg11}.

\begin{figure}
\includegraphics[width=8cm]{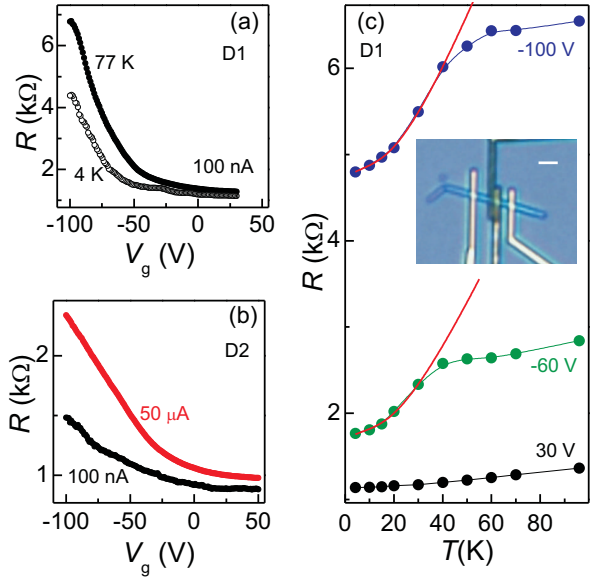}
\vspace{-2mm}
\caption{(a) Resistance $R$ as a function of the gate voltage $V_{g}$ at T = 77 K (solid symbols) and at T = 4.2 K (empty symbols), with applied current $I$ = 100 nA, for device D1.(b) $R$ vs $V_g$ for device D2 at $T= 77$ K, $I = 100$ nA (black symbols), $I = 50$ $\mu$A (red symbols). (c)  $R$ vs temperature $T$ for device D1 at $V_{g}$ (+30 V, -60 V, -100 V). The red lines are best fits to Eq. (1). The inset shows an optical image of a cleaved crystal contacted with four electrodes.} \label{Fig1}
\end{figure}

When $\mu$ lies inside the gap, the conductance $G=1/R$ splits in surface $G^S$ and bulk $G^B$ contributions, $G= G^S+G^B$, where $G^B$ results from carriers excitation to the conduction or valence bands, which is more significant for $T \gtrsim 100$ K \cite{ong11,fuhrer12}. Hence, the strong temperature dependence for negative $V_g$ at $T < 50 $ K suggests an activated process at the surface mediated by inelastic scattering involving a single or multiple phonon modes. Our data can be indeed fitted (red lines on Fig. 1c) with an equation that only includes a single Bose-Einstein term, thus a single phonon mode, as follows:
\begin{equation}
R^S(V_{g},T)=R_{0}(V_{g}) + A (V_{g})\times T + B (V_{g}) \times \left( \frac{1}{e^{\hbar\Omega/k_{B}T}-1} \right ),\label{RvsT}
\end{equation}
\noindent where the first term $R_{0}(V_{g})$ accounts for low temperature (residual) resistance due to scattering on static impurities or defects, while the second term is associated to acoustic phonons, for which a linear resistance with temperature has been predicted \cite{egger11} and recently observed \cite{fuhrer12}.

In Eq. (1), $A$ and $B$ are fitting parameters, where $A$ is independent on $V_g$ for carriers with a linear dispersion relation, that is, for Dirac fermions at the surface \cite{fuhrer08}. The best fit gives $7.7 \pm 1$ meV which corresponds to the energy of the dominant surface phonon branch observed by helium-beam surface scattering \cite{batanouny12} and optical methods \cite{basov10,dipietro13}. It could also be due to scattering off bulk optical phonons projected into the surface \cite{batanouny11}. The fit however is not enough proof to demonstrate that the main scattering mechanism originates from an optical phonon branch. A power-law dependence expected in some cases for acoustic phonons scattering can also fit the results, and therefore further experiments are necessary \cite{SI}.

An activated response similar to that in Fig. 1c was found in graphene on a SiO$_2$ substrate \cite{fuhrer08}. It was argued to originate from remote interfacial phonon scattering by surface optical phonon modes in SiO$_2$. Because the lowest energy of the relevant modes exceeds 50 meV, much larger than that in  Bi$_{2}$Se$_{3}$, the activated behavior was only observed at $T > 200$ K. Additionally, the same nonlinear $T$-dependence of $R$ was noted in Ca-doped Bi$_{2}$Se$_{3}$ crystals and its onset with $V_g$ was interpreted as $\mu$ reaching the lower edge of the conduction band \cite{ong11}, however the activated nature of the effect was not discussed \cite{SI}.

Electrical transport experiments in graphene have shown that the current tends to saturate as the bias voltage is increased above $\sim 200$ meV \cite{meric08, barreiro09,Foa2006}. The saturation has been attributed to scattering of electrons by optical phonons, either activated at the SiO$_2$ surface \cite{meric08,scharf13} or intrinsic to graphene \cite{barreiro09}. If the activation energy observed in Fig. 1c was due to an optical phonon mode, the current in Bi$_{2}$Se$_{3}$ should also tend to saturate with $V$. Such saturation behavior is expected for scattering of electrons by optical phonons but not for acoustic phonons. Additionally, the saturation current should be weakly temperature dependent.

A first indication of sublinear $I$-$V$ response is observed in Fig. 1b, where an increase in $R$ is evident when a large current $I= 50$ $\mu$A is applied. The fact that the relative change of $R$ is much larger when $\mu$ is shifted towards the bulk gap \cite{ong11} supports that the main phenomenon is surface related.

Figures 2 and 3 show our main results. Figure 2 displays the $I$-$V$ characteristics for different gate voltages $V_g$ at 4.2 K. The current at large negative $V_g$ presents a clear tendency for saturation at $V < 50$ mV, with a smaller saturation current $I_S$ for larger absolute values of $V_g$. This behavior is consistent with optical-phonon scattering and resembles the behavior observed in graphene, where the current saturation becomes gate dependent and decreases when approaching the Dirac point \cite{meric08, barreiro09}. However, $I(V)$ also shows a kink at $V \sim 40$ mV. As discussed below, this feature is likely to be associated to carrier excitations into the conduction and/or valence bands.

\begin{figure}
\includegraphics[width=8cm]{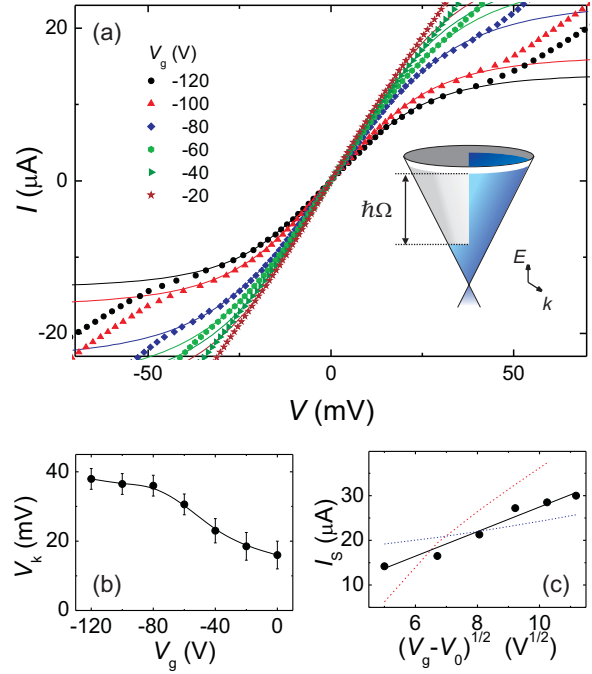}
\vspace{-2mm}
\caption{(a) Current voltage characteristics $I$ vs $V$ at different $V_g$ (from -20 V to -120 V, as labelled) for device D2. The lines are best fit to a current saturation model (see text). (b) Position of the voltage kink, $V_k$, as a function of $V_g$. The line is a guide to the eye. (c) Saturation current $I_S$ vs $V_g$ (black circles). The lines are best fits to Eq. (2) with both $V_0$ and $\hbar \Omega$ as fitting parameters, resulting in $V_0$ = -145 V and $\hbar \Omega$ = 8 meV (solid black). Best fits for fixed $\hbar \Omega =$ 5 meV (dotted blue) and $\hbar \Omega =$ 12 meV (dotted red) are shown for comparison.} \label{Fig2}
\end{figure}

Because the saturation is not complete, we approximate $I$ by a saturation model for further analysis. Having compared the fit of our data with different analytical expressions \cite{meric08,canali75,sze85,Pop10}, we found that, as in graphene \cite{Pop10}, the best results are obtained using $I(V)= G_0 V /[1+(G_0V/I_S)^{\beta}]^{1/\beta}$, where $G_0$ is the gate-dependent conductance at low $V$, and $\beta$ and the saturation current $I_S$ are fitting parameters. We fitted the data up to the position of the kink $V_k$, which is determined by the minimum in $dI/dV$ \cite{SI}. The gate-dependence of $V_k(V_g)$ is shown in Fig. 2b. The above equation with $\beta =2.5$ gives the best fit, regardless of the value of $V_g$ (lines in Fig. 2a); the extracted values of $I_S$ are shown in Fig. 2c.

In order to model $I_S$, we note that, for instantaneous phonon emission, a steady-state population is established in which right moving electrons are populated to a higher energy $\hbar\Omega$ than left moving ones \cite{yao00,barreiro09} (see inset of Fig. 2a). $I_S$ can then be estimated by integrating the electrons velocity over the steady state population, which for Dirac fermions and two coupled surfaces results in \cite{barreiro09,Eq1}:
\begin{equation}
I_S \approx \frac{ W}{\pi^{\frac{3}{2}}\hbar} \hbar\Omega \sqrt{e C_g (V_g-V_0)},\label{IsvsVg}
\end{equation}

\noindent
where $W$ is the width of the sample and $C_g$, the backgate capacitance. $V_0$ is the gate voltage at which $\mu$ is tuned at the Dirac point.

Even though we cannot access $V_0$, because the threshold for breakdown of the SiO$_2$ dielectric is reached first, direct comparison with results in Ca-doped Bi$_{2}$Se$_{3}$ crystals \cite{ong11} together with the incipient rounding of the $R(V_g)$ response (Figs. 1b and 1c) suggest that $V_0$ is within a few tens of volts beyond the breakdown. Indeed, Eq. (2) gives a good fit to $I_S$ vs $V_g$ with $V_0 \approx -145$ V (Fig. 2c). From the fit, we obtain $\hbar \Omega = 8.1 \pm 1.5$ meV, which is in remarkable agreement with the activation energy estimated from the data in Fig. 1. Additionally, Fig. 2c shows the large discrepancy between the experimental data and the theoretical $I_{S}$ using Eq. (2) for $\hbar \Omega$ equal to 5 meV and 12 meV, which further yield unrealistic values of $V_0$ (-245 V and -120 V, respectively). This demonstrates that $\hbar \Omega$ must be in a very restricted range of energies.

Figure 3 shows the conductance $G(V)$ at specific gate voltages. At high positive $V_{g}$, $G$ presents a smooth decrease with $V$, which is typical of a bulk metal in which the dominant scattering mechanism is disorder. At large negative $V_{g}$, when activation effects in Fig. 1 become evident, $G$ develops a plateau-like feature under the action of a weak electric field. There, the only contribution to the resistance comes from elastic scattering induced by static disorder. The increase of the bias voltage leads to an energy gain by the propagating electrons, which eventually reaches the energy threshold that allows for an electron-phonon scattering event (such as phonon emission). A new inelastic transport length scale then enters into play and brings about an additional contribution to the resistance, which adds up to the elastic one, in virtue of the Matthiessen rule. At fixed gate voltage (or Fermi level position), the elastic part is given by the measured resistance at low-temperature and zero-bias, whereas the inelastic contribution is bias-dependent. The onset for phonon scattering occurs at about 10 mV (pinpointed by vertical lines in Figs. 3 and 4), while in this crystal the suppression of the conductance becomes significant only beyond 20-30 mV.

\begin{figure}[h]

\includegraphics[width=9cm]{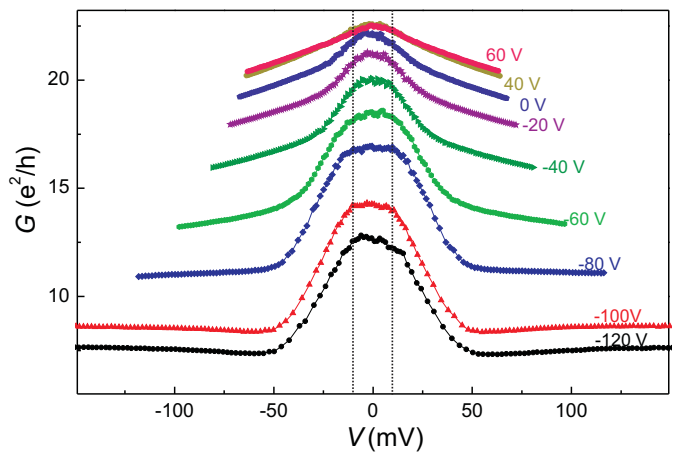}
\vspace{-2mm}
\caption{Conductance ($G=I/V$) vs $V$ at fixed gate voltages $V_{g}$ (from -120 V to +60 V) and at T = 4.2 K for device D2.} \label{Fig3}
\end{figure}

We now focus on the origin of the kink in the $I$-$V$ characteristics (Fig. 2a). The position of the kink $V_k$, which is given in Fig. 2b, is weakly dependent on $V_g$ for $V_g < -80$ V, but drops quickly as $V_g$ approaches zero, becoming undetectable for positive $V_g$, which coincides with $eV_k \sim \hbar \Omega$.

The kink can be due to excitations of carriers into the conduction and/or valence bands \cite{ong11, fuhrer12, wu13}. A similar gate-driven transformation to a conductor with strong temperature dependence (Fig. 1c) was reported in Ref. \onlinecite{ong11} and associated to the opening of an effective bulk gap $\Delta \sim 50$ meV that results from gap narrowing due to band bending (the intrinsic bulk gap in Bi$_2$Se$_3$ is about 300 meV). At low temperatures, the presence of such gap is essential for observing signatures of surface states. However, at high enough temperatures, thermal excitation of carriers from remnant electron pockets to the conduction band can enhance the bulk transport contribution \cite{ong11,SI}. The transfer of electrons from the electron pockets to the conduction band can also be driven by high electric fields, which for $V=50$ meV exceed 1 kV/cm. Such transitions are known to occur in GaAs between conduction valleys \cite{sze85}. This explanation is further supported by the increase of $V_k$ as the chemical potential is moved deeper into the gap; however the weak dependence for $V_g < -80$ V remains unexplained. Thermal activation of a bulk channel was also observed in Ref. \onlinecite{fuhrer12}, but ascribed to the activation of electrons from the bulk valence band to the surface band. It is plausible that the overall response of $V_k$ results from the combination of these two phenomena, and thus $V_g \sim -80$ V may signal the onset of electron transfer from the valence band that prevents $V_k$ from further increasing.

Figure 4 displays $G(V)$ at $V_{g}=-100$ V for $T$ between 4.2 K and 70 K. $G$ is strongly temperature dependent for $|V| < 50$ mV. Over this $V$ range and for $T < 40$ K, $G$ presents the same activated behavior that is observed at low $V$, with a characteristic energy of about 8 meV (inset). In contrast, $G$ is largely temperature independent for $V> 50$ mV. This is typical of a disordered metal, a behavior that was found when $\mu$ lies in the bulk conduction band (Fig. 1c), which is in agreement with the hypothesis of electron transfer into bulk channels. Finally, the observation of $I_S$ being nearly independent of $T$ despite significant change in $G$, further supports the origin of the saturation as being related to a single optical phonon branch \cite{SI}.

\begin{figure}[h]
\includegraphics[width=8.5cm]{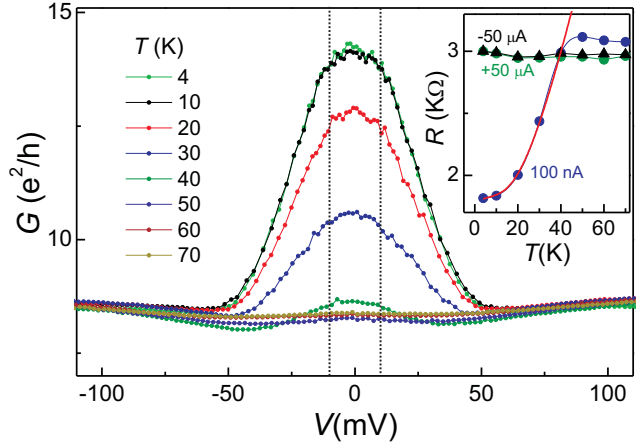}
\vspace{-2mm}
\caption{$G$ vs $V$ at fixed $T$ (from 4 K to 70 K) for  $V_{g}$= -100 V. Inset $R$ vs $T$ inferred from data shown in the main panel at $I$ = 100 nA, +50 $\mu A$ and -50 $\mu A$. The red line is the best fit to Eq. (1).} \label{Fig4}
\end{figure}

In conclusion, our results show compelling evidence of a strong \textit{e-ph} coupling involving an optical phonon mode with $\hbar \Omega \approx 8$ meV that mediates inelastic scattering. The thermally activated behavior of the Bi$_{2}$Se$_{3}$ resistance at low temperatures, the voltage at which inelastic scattering emerges, and the magnitude of the saturation current at high voltage and its independence of temperature are all consistent with this interpretation. This is reinforced by helium scattering experiments, which identified a main phonon mode at 6-8 meV and the absence of acoustic Rayleigh phonons \cite{batanouny11,batanouny12}, and theoretical calculations that show a bulk optical phonon that projects into the surface Brillouin zone in the same energy range \cite{batanouny11}. The transport experiments carried out here are in a less controlled environment than those in Refs. \cite{batanouny11,batanouny12}. However, the surface states in Bi$_{2}$Se$_{3}$ have proven to be robust to processing and, in particular, the relevant surface phonon mode was observed even after exposure to air \cite{basov10}.

We acknowledge the support from the European Research Council (ERC Grant agreement 308023 SPINBOUND) MINECO (MAT2010-18065 and MAT2012-33911, RYC-2011-08319) and AGAUR (Beatriu de Pin\'{o}s program).


\begin{thebibliography}{99}
\bibitem{Eq1} In deriving the Eq. (2), we considered the second expression of Eq. (1) from Ref. \cite{barreiro09}; due to the low energy of the relevant phonon modes we assumed $(\hbar\Omega)^{2}/4$ $\ll$  $(C_{g} V_{g} (\hbar v_{F})^{2} \pi)/e$, a condition that is later verified by the fitting results.
\end{thebibliography}
\end{document}